\documentclass[12pt]{article}
\usepackage{graphicx}
\usepackage{wrapft}
\usepackage{wrapfig}
\usepackage{cite}
\usepackage{amsmath}
\usepackage{amssymb}
\usepackage{amsfonts}
\usepackage{physics}
\usepackage{ulem}
\usepackage{color}
\numberwithin{equation}{section}
\def\be{\begin{eqnarray}}
\def\ee{\end{eqnarray}}
\def\eq{\label}

\def\abstract#1{\vskip 7mm 
\begin{center}{\large Abstract}\par \bigskip
\begin{minipage}[c]{12cm}
\small #1
\end{minipage}
\end{center}
}
\def\title#1{\begin{center}{\Large\bf #1}\end{center}}
\def\author#1{\vskip 5mm \begin{center}{#1}\end{center}}
\def\address#1{\begin{center}{\it #1}\end{center}}
\newcommand{\bfr}{\begin{flushright}}
\newcommand{\efr}{\end{flushright}}
\begin{document}
\title{Gauge-Theoretical Method in Solving Zero-curvature Equations III 
\\[1ex]\large{---Gauge Theoretical method and the B\"{a}cklund Transformations for Solitons}---}
\vspace{2cm}
\author{
T.~Azuma
\footnote{azuma@dokyo.ac.jp}
}
\address{Dokkyo University\\
1-1 Gakuencho, Soka, Saitama 340-0042, Japan 
}
\author{
T.~Koikawa
\footnote{koikawa@otsuma.ac.jp}
}
\address{Institute of Human Culture Studies, Otsuma Women's University\\
12 Sanban-cho, Chiyoda-ku, Tokyo 102-8357, Japan}
\hspace{0.54cm}
\vspace{2cm}
\abstract{In soliton theory, both the gauge-theoretical method and the B\"acklund transformation yield soliton equations from the compatibility condition of a pair of linear equations. Therefore, it is necessary to clarify the similarities and differences between these two methods. The B\"acklund transformation defines a transformation from one soliton solution to another. In particular, restricting the transformation to solutions with different soliton numbers yields interesting insights. Using several examples, we demonstrate how soliton solutions with different soliton numbers are constructed.
}
\thispagestyle{empty}
\newpage
\section{Introduction}

In the previous papers\cite{AK1,AK2}, we demonstrated that numerous soliton equations can be obtained via the compatibility condition of a pair of first-order differential equations. These first-order equations are expressed by the vanishing of the covariant derivatives of a wave function with respect to space and time in the gauge theory. In this framework, a wave function and gauge fields appear naturally. The compatibility condition of the pair of covariant derivatives leads to the zero-curvature equation of the gauge fields, which corresponds to the soliton equations.

Since soliton equations are higher-order partial differential equations (typically second-order or higher), it is computationally easier to tackle the associated first-order equations. Although the idea that soliton equations can be derived from the compatibility condition of a pair of first-order differential equations is inherent in the concept of the B\"acklund (BT) transformation, it is necessary to distinguish this analogous concept from the gauge-theoretical approach, where gauge fields and a wave function are explicitly introduced. In the B\"acklund transformation approach, a field satisfying the soliton equation appears. We shall discuss this in Section 4 to clarify the differences and similarities between this approach and the gauge-theoretical approach.

This paper is constructed as follows. In the following section, we present the gauge theoretical approach to the soliton equations. In Section 3, we review Hirota's method and the procedure for obtaining a 2-soliton solution using this method. In Section 4, we demonstrate the BT and the method for obtaining a two-soliton solution by using this transformation, while clarifying its similarities and differences from the gauge theoretical approach. The last section is devoted to the summary and discussion.   

\section{Gauge Theoretical Construction of Soliton Equations}
Various soliton equations \cite{miura,GGKM,ZakShab1,ZakShab2,Wadachi1,Wadachi2,akns,aks} and the Einstein equation \cite{BZ1,BZ2,BV} can be derived from the compatibility condition of a pair of first-order differential equations that correspond to the covariant derivatives in gauge theory. In this section, we present the procedure for this gauge theoretical construction of soliton equations. 

We begin with two linear differential equations in the coordinates $x$ and $t$, given by
\be
\left(\partial_x-A_x(\lambda,x,t)\right)\psi(\lambda,x,t)&=&0,\eq{linear1}\\
\left(\partial_t-A_t(\lambda,x,t)\right)\psi(\lambda,x,t)&=&0,\eq{linear2}
\ee 
where $A_x$ and $A_t$ are matrix-valued gauge potentials, $\psi(\lambda,x,t)$ is a complex matrix function, and $\lambda$ is a spectral parameter. In these equations, $\partial_x-A_x$ and $\partial_t-A_t$ are called covariant derivatives in the gauge theory. The compatibility condition of Eqs. (\ref{linear1}) and (\ref{linear2}),
\be
\left[\partial_x-A_x,\partial_t-A_t\right]\psi=0
\ee
leads to
\be
\left(\partial_xA_t-\partial_tA_x+[A_t, A_x]\right)\psi=0.\eq{zcpsi}
\ee
For a nontrivial $\psi$ to exist, we obtain the zero-curvature equation:
\be
\partial_xA_t-\partial_tA_x+[A_t, A_x]=0.\eq{zerocurvature}
\ee

Next, we demonstrate how the zero-curvature equation (\ref{zerocurvature}) derives soliton equations, using the KdV equation and the sine-Gordon equation as examples. 

For the potentials $A_x$ and $A_t$, if we choose 
\be
A_x&=&i\lambda\mqty(1 & 0 \\ 0 & -1)+\mqty(0 & 1 \\ -u & 0),
\ee
and
\be
A_t&=&4i\lambda^3\mqty(1 & 0 \\ 0 & -1)+4\lambda^2\mqty(0 & 1 \\ -u & 0)\cr
&+&2i\lambda\mqty(-u & 0 \\ -u,_x & u)+\mqty(u,_x & -2u \\ 2u^2+u,_{xx} & -u,_x),
\ee
we then get from Eq.({\ref{zerocurvature})
\be
\mqty(0 & 0 \\ u,_t+6uu,_x+u,_{xxx} & 0)=0,
\ee
which gives the KdV equation:
\be
u,_t+6uu,_x+u,_{xxx}=0.
\ee

Alternatively, assuming $(x,t)$ to be a pair of null coordinates and choosing
\be
A_x&=&i\lambda\mqty(1 & 0 \\ 0 & -1)+\frac{i}{2}\mqty(0 & u,_x \\ u,_x & 0),
\ee
and
\be
A_t&=&\frac{1}{4i\lambda}\mqty(\cos u & -i\sin u \\ i\sin u & -\cos u),
\ee
we obtain from Eq.({\ref{zerocurvature})
\be
\mqty(0 & u,_{xt}-\sin u \\ u,_{xt}-\sin u & 0)=0,
\ee
which is the sine-Gordon equation in the null coordinates:
\be
u,_{xt}-\sin u=0.
\ee

The gauge transformation given by
\be
\psi &\to& \psi'=G\psi,\\
A_x &\to& A'_x=G A_xG^{-1}-(\partial_xG)G^{-1},\\
A_t &\to& A'_t=G A_tG^{-1}-(\partial_tG)G^{-1},
\ee
where $G=G(\lambda,x,t)$ is a $N\times N$ matrix, leads to a new zero-curvature equation:
\be
\partial_xA'_t-\partial_tA'_x+[A'_t, A'_x]=0.\eq{zerocurvature1}
\ee
Therefore, if we have a solution to a soliton equation satisfied by the potentials $A_x$ and $A_t$, we then get  a new soliton solution from the gauge transformed potentials $A'_x$ and $A'_t$

The above procedure can be applied to the stationary axisymmetric vacuum Einstein equation\cite{BZ2} and the electrovac or magnetovac Einstein equations\cite{AK1,AK2} by extending the differential operators $\partial_x$ and $\partial_t$ to commuting operators $\hat D_1$ and $\hat D_2$.  These operators consist of linear combinations of differentiations with respect to the canonical cylindrical coordinates $(\rho,z)$ and the spectral parameter $\lambda$. The zero-curvature equation, derived from the compatibility condition, yields a zero-curvature equation for the potentials with $\lambda=0$ and  part of the Einstein equations. Previous papers\cite{AK1,AK2}  demonstrate the application of the gauge-theoretical method to the zero-curvature equation in static Einstein-Maxwell equations involving magnetic charge.

\section{Hirota's Bilinear Form for Soliton Equations}
One of the most powerful tools for solving the soliton equations is Hirota's bilinear method. In the following section, we present the B\"{a}cklund transformation(BT) for soliton equations and demonstrate
 how it is applied to obtaining soliton solutions. The procedure for obtaining a two-soliton solution using Hirota's bilinear equation is then compared to that using BT. 

In this section we first demonstrate how to obtain soliton solutions by using Hirota's bilinear form for the KdV equation. We note that the most difficult part of this method is deriving the bilinear form equation from the original partial differential equation. Following the discussion of the KdV equation, we show the B\"{a}cklund transformation within the bilinear formalism.

Hirota introduced a $D$-operator\cite{hiro1} which is defined by
\be
D^m_xD^n_y f\cdot g=\lim_{{x'}\to x}\partial^m_x\partial^n_{x'} f(x)g(x').
\ee
\subsection{Bilinear form of the KdV equation}
We show how to obtain soliton solutions in Hirota'a bilinear formalism, exemplified by the KdV equation. The bilinear KdV equation is given by
\be
P(D_x,D_t)f\cdot f=0,
\ee
where
\be
P(D_x,D_t)=D_x(D_t+D_x^3).\eq{blkdv}
\ee
We expand $f$ in powers of $\epsilon$ as
\be
f=f_0+\epsilon f_1+\epsilon^2f_2+\cdots.
\ee
At each power of $\epsilon$, we obtain
\be
&&O(\epsilon^1):P(D_x,D_t)(f_1\cdot 1+1\cdot f_1)=0,\eq{lineq}\\
&&O(\epsilon^2):P(D_x,D_t)(f_2\cdot 1+1\cdot f_2+f_1\cdot f_1)=0,\eq{lineq2}\\
&&\cdots. \nonumber
\ee

For the 1-soliton solution, the expansion is truncated as 
\be
f=1+\epsilon f_1,
\ee
with
\be
f_1&=&e^{\eta},
\ee
where $\eta=k x-\omega t+\eta_0$, and we assume that $f_i=0$ for $i \geq 2$.
Then Eq.(\ref{lineq}) becomes as
\be
P(D_x,D_t)(e^\eta\cdot 1+1\cdot e^\eta)=0.
\ee
This equation is solved as
\be
P(D_x,D_t)(e^\eta\cdot 1+1\cdot e^\eta)=2P(\partial_x,\partial_t)e^\eta=2k(-\omega+k^3))e^\eta=0.
\ee
This equation yields the dispersion relation relating $\omega$ with $k$, meaning they are no longer independent. Consequently, the 1-soliton solution is given by
\be
f=1+e^{kx-k^3t+\eta_0},
\ee
where we set $\epsilon=1$.

Next, to discuss the 2-soliton solution, we make use of two 1-soliton solutions $e^{\eta_1}$ and $e^{\eta_2}$, and assume that the 2-soliton solution tales the following form:
\be
f=1+\epsilon f_1+\epsilon^2 f_2,
\ee
with
\be
f_1&=&e^{\eta_1}+e^{\eta_2},\\
f_2&=&a_{12}e^{\eta_1+\eta_2},
\ee
where
\be
\eta_i=k_ix-k_i^3t+\eta^i_0.~~(i=1,2)
\ee
Here $a_{12}$ is a constant to be determined. Substituting these equations into Eq.(\ref{lineq2}), we obtain
\be
&&P(D_x,D_t)\left(a_{12}(e^{\eta_1+\eta_2}\cdot 1+1\cdot e^{\eta_1+\eta_2})+e^{\eta_1}\cdot e^{\eta_2}+e^{\eta_2}\cdot e^{\eta_1}\right)\cr
&=&2\left(P(\partial_x,\partial_t)(a_{12}e^{\eta_1+\eta_2})+P(D_x,D_t)e^{\eta_1}\cdot e^{\eta_2}\right)\cr
&=&2\left(a_{12}P(k_1+k_2,-(\omega_1+\omega_2)+P(k_1-k_2,\omega_1-\omega_2)\right)e^{\eta_1+\eta_2}\cr
&=&0.
\ee
Solving this for $a_{12}$ yields
\be
a_{12}=-\frac{P(k_1-k_2,\omega_1-\omega_2)}{P(k_1+k_2,-(\omega_1+\omega_2)}.
\ee
Using Eq.(\ref{blkdv}), this is computed as
\be
a_{12}=\frac{(k_1-k_2)^2}{(k_1+k_2)^2}.
\ee
We have thus obtained the 2-soliton solutions to the KdV equation.

In the next section, the B\"{a}cklund transformations for various soliton equations  are shown. We show how we obtain the 2-soliton solution via the B\"{a}cklund transformations. The comparison with the method of obtaining the 2-soliton solution via the bilinear form and the B\"{a}cklund transformations is useful in unerstanding the method of getting the solutions with higher soliton numbers.

Last of this section, we present the B\"{a}cklund transformations for the KdV equation via the bilinear form\cite{hiro2}. Let $u$ and $u'$ be different solutions to the KdV equation, and introduce two different solutions $f$ and $g$ to the Hirota's Bilinear form equations:
\be
u=2(\ln f)_{xx}, \\
u'=2(\ln g)_{xx}.
\ee
Then $f$ and $g$ both satisfy the  Hirota's bilinear equations for the KdV equations:
\be
(D_x^3+D_t)f\cdot f&=&0,\\
(D_x^3+D_t)g\cdot g&=&0.
\ee
The B\"{a}cklund transforamtions for the KdV equation consist of a coupled equations
\be
(D_x^2-\lambda)g\cdot f&=&0,\\
(D_x^3+D_t-3\lambda D_x)g\cdot f&=&0,
\ee
where $\lambda$ is a parameter.

\section{B\"{a}cklund transformation}
A B\"{a}cklund transformation(BT) transforms a nonlinear partial differential equation (PDE) into another PDE. Since BT relates PDEs and their solutions, it is widely used to generate new solutions to nonlinear PDEs. Therefore BT is used to generate new solutions to nonlinear PDEs. BTs typically consist of a pair of two first-order PDEs involving two functions. If these two functions satisfy the PDEs separately, they are related by the BT. One of the interesting feature of BT is that the compatibility condition leads to PDEs, which should be compared to our gauge theoretical construction of soliton equations\cite{AK1,AK2}, where we also derive PDEs as a compatibility condition. However there is a large difference between these two methods. PDEs in our method are described by the zero-curvature equation of gauge fields. On the other hand, there is no gauge fields in BT formalism which we discuss in this section.

In this section, we first illustrate BT using for the potential KdV equation. We then present the BT for the Liouville equation, and finally demonstrate the BT for the sine-Gordon equation, showing how to obtain the 1-soliton and 2-soliton solutions.

\subsection{B\"{a}cklund Transformation for the potential KdV equation}
Suppose that the $u$ satisfies the KdV equation: 
\be
u_t+6uu_x+u_{xxx}=0.
\ee
We introduce $v$ by
\be
v_x=u.
\ee
Then $v$ satisfies 
\be
v_t+3v_x^2+v_{xxx}=0.
\ee
The equation that $v$ governs is called a potential KdV(p-KdV) equation. In this subsection, we discuss the {B\"{a}cklund transformation for the potential KdV equation and show how it works to obtain a soliton solution.

The B\"{a}cklund transformations(BTs) for the potential KdV \cite{wahesta} fields $w$ and $w'$ are given by
\be
(w+w')_x&=&\lambda-\frac{1}{2}(w-w')^2,\eq{plus1}\\
(w-w')_t&=&-(w-w')_{xxx}-3(w_x^2-{w'}_x^2).\eq{minus2}
\ee
In this subsection, we first show that the compatibility condition of the BTs lead to the p-KdV equation. Next we show how we obtain an 1-soliton solution from a trivial solution by using the BTs. 

The $t$}-derivative of Eq.(\ref{plus1}) is given by
\be
(w+w')_{xt}&=&-(w-w')\partial_t(w-w').\eq{cmp1}
\ee
The $x$-derivative of Eq.(\ref{minus2}) is obtained as
\be
(w-w')_{tx}&=&-(w-w')_{xxxx}-3\partial_x(w_x^2-{w'_x}^2).\eq{cmp2}
\ee
By substituting Eq.(\ref{minus2}) into the rhs of Eq.(\ref{cmp1}) we obtain
\be
(w+w')_{xt}&=&(w-w')(w-w')_{xxx}+3(w-w')(w_x-w'_x)(w_x+w'_x).\ \ \ \ \eq{cmp3}
\ee
In order to rewrite the rhs of this equation, we prepare the following equations derived from Eq.(\ref{plus1}):
\be
(w+w')_{xx}&=&-(w-w')(w-w')_x,\\
(w+w')_{xxxx}&=&-3(w-w')_x(w-w')_{xx}-(w-w')(w-w')_{xxx}.\ 
\ee
Using these equations, the rhs of Eq.(\ref{cmp3}) becomes
\be
&&(w-w')(w-w')_{xxx}+3(w-w')(w_x-w'_x)(w_x+w'_x)\cr
&=&-\frac{3}{2}\partial_x(w_x-w'_x)^2-(w_x+w'_x)_{xxxxx}-3(w_x+w'_x)(w+w')_{xx}\cr
&=&-\frac{3}{2}\partial_x(w_x-w'_x)^2-(w_x+w'_x)_{xxxxx}-\frac{3}{2}\partial_x(w_x+w_x')^2\cr
&=&-\frac{3}{2}\partial_x(w_x-w'_x)^2-(w_x+w'_x)_{xxxxx}-\frac{3}{2}\partial_x(w_x+w_x')^2\cr
&=&-3\partial_x(w_x^2+{w'_x}^2)^2-\partial_x(w+w')_{xxx}.
\ee
Then Eq.(\ref{cmp1}) reads
\be
\partial_x(w+w')_t=-\partial_x\left(3(w_x^2+{w'_x}^2)^2+\partial_x(w+w')_{xxx}\right).
\ee 
Finally we  obtain
\be
(w+w')_t=-(w+w')_{xxx}-3(w_x^2+{w'_x}^2)^2\eq{plus_t}+C,
\ee
where $C$ is an integral constant and hereafter we set $C=0$. The sum of Eq.(\ref{minus2}}) and Eq.(\ref{plus_t}) leads to
\be
w_t+3w_x^2+w_{xxx}=0.
\ee
and the subtraction
\be
w'_t+3w_x'^2+w'_{xxx}=0.
\ee

Next we shall derive a 1-soliton solution from a trivial solution. We start with setting  $w'=0$ for the trivial solution. Then the BT reads
\be
w_x&=&\lambda-\frac{1}{2}w^2,\eq{1sol1}\\
w_t&=&-w_{xxx}-3w_x^2.\eq{1sol2}
\ee
The $x$-part Eq.(\ref{1sol1}) is integrated as
\be
\int\frac{1}{a^2-w^2}dw=\frac{1}{2}\int dx\eq{afterint},
\ee
where we set $a^2=2\lambda$. Then we obtain
\be
\ln\left|\frac{a+w}{a-w}\right|=ax+C(t).
\ee
Assuming $a<|w|$, the absolute symbol is removed and this yields
\be
\ln\frac{a+w}{a-w}=ax+C(t).\eq{adopted}
\ee
Solving this relation for $w$, we obtain
\be
w=a\tanh(\frac{a}{2}x+C(t)).
\ee
Next we integrate Eq.(\ref{1sol2}) and determine $C(t)$. This leads to
\be
C_t(t)=-\frac{a^3}{2}.
\ee
Thus the B\"{a}cklund transformation  brings about a non-trivial solution:
\be
w=\sqrt{2\lambda}\tanh\left(\sqrt{\frac{2\lambda}{2}}(x-2\lambda t)+c_0)\right).
\ee
Starting with a trivial solution we obtain 1-soliton solution. It is possible to start with 1-soliton solution to obtain 2-soliton solution and more.
\subsection{B\"{a}cklund Transformation for the Liouville equation}
Suppose $u=u(x,y)$ and $v=v(x,y)$ are related by the following B\"acklund transformation  \cite{rosa, baby}
\be
v_x=u_x+2a\exp\left(\frac{u+v}{2}\right),\\
v_y=-u_y-\frac{1}{a}\exp\left(\frac{u-v}{2}\right).
\ee
We rewrite these equations to
\be
(u-v)_x&=&-2a\exp\left(\frac{u+v}{2}\right),\eq{umvx}\\
(u+v)_y&=&-\frac{1}{a}\exp\left(\frac{u-v}{2}\right)\eq{upvy}.
\ee
{
Taking the derivative of Eq.(\ref{umvx}) with respect to $y$ and of Eq.(\ref{upvy}) with respect to $x$ gives}
\be
(u-v)_{xy}&=&-2a\exp\left(\frac{u+v}{2}\right)\frac{1}{2}(u+v)_y\cr
&=&-a\exp\left(\frac{u+v}{2}\right)\left(-\frac{1}{a}\right)\exp\left(\frac{u-v}{2}\right)\cr
&=&\exp u.\\
(u+v)_{yx}&=&-\frac{1}{a}\exp\left(\frac{u-v}{2}\right)\frac{1}{2}(u-v)_x\cr
&=&-\frac{1}{a}\exp\left(\frac{u-v}{2}\right)\frac{1}{2}(-2a)\exp\left(\frac{u+v}{2}\right)\cr
&=&\exp u.
\ee
Addition of these equations leads to
\be
u_{xy}=\exp u.
\ee
Subtaction leads to
\be
v_{xy}=0.
\ee

\subsection{B\"{a}cklund Transformation for the sine-Gordon equation}
In the previous section, we showed how we obtain a 2-soliton solution in the Hirotas bilinear form, where the 2-soliton solution is constructed by two different 1-soliton solutions. In this section, we discuss how we obtain a 2-soliton solution in the B\"{a}cklund transformation formalism by exemplifying the sine-Gordon equation\cite{lamb}.

The B\"{a}cklund transformation(BT) for the sine-Gordon equation is given by
\be
\frac{1}{2}(u-v)_x=a\sin\left(\frac{u+v}{2}\right),\eq{SGBT1}\\
\frac{1}{2}(u+v)_t=\frac{1}{a}\sin\left(\frac{u-v}{2}\right).\eq{SGBT2}
\ee
By differentiating the first equation with respect to $t$, we obtain
\be
\frac{1}{2}(u-v)_{xt}=\cos\left(\frac{u+v}{2}\right)\sin\left(\frac{u-v}{2}\right).
\ee
By differentiating the second equation with respect to $x$, we obtain
\be
\frac{1}{2}(u+v)_{xt}=\cos\left(\frac{u-v}{2}\right)\sin\left(\frac{u+v}{2}\right).
\ee
The addition and subtraction of these equations lead to
\be
u_{xt}&=&\sin u,\\
v_{xt}&=&\sin v.
\ee
We find both $u$ and $v$ satisfy the sine-Gordon equation. Such BT is called an invariant B\"{a}cklund transformation or an auto-B\"{a}cklund transformation.

Next we obtain the 1-soliton solution via BT. Setting $v=0$ in Eqs.(\ref{SGBT1}) and (\ref{SGBT2}), we obtain
\be
\frac{1}{2}u_x&=&a\sin\left(\frac{u}{2}\right),\eq{pdx}\\
\frac{1}{2}u_t&=&\frac{1}{a}\sin\left(\frac{u}{2}\right).\eq{pdt}
\ee
Integrating Eq.(\ref{pdx}) yields:
\be
\ln|\tan\left(\frac{u}{4}\right)|=ax+b(t),
\ee
where we write the integral constant as $b(t)$ because it may depend on time $t$. This can be rewritten as
\be
u=4\arctan\left(Ce^{ax+b(t)}\right).\eq{sola}
\ee
Similarly, integrating Eq.(\ref{pdt}) with respect to $t$ gives:
\be
u=4\arctan\left(Ce^{\frac{1}{a}t+B(x)}\right).
\ee
Comparing these two equations, we find $B(x)=ax$ and $b(t)=\frac{1}{a}t$. 
Then the 1-soliton solution becomes
\be
u=4\arctan \left(Ce^{ax+\frac{1}{a}t} \right).
\ee

Next we compute the 2-soliton solution for the sine-Gordon equation. A nonlinear superposition principle of the B\"acklund transformations states that successive applications of two different B\"acklund transformations yield the same result.

Let $u_0$ be the trivial solution ($u_0=0$). Suppose $u_1$ is generated from $u_0$ with $a_1$, and $u_2$ is generated with parameter $a_2:$
\be
\frac{1}{2}(u_1-u_0)_x&=&a_1\sin\left(\frac{u_1+u_0}{2}\right),\\ \eq{SGBT11}
\frac{1}{2}(u_2-u_0)_x&=&a_2\sin\left(\frac{u_2+u_0}{2}\right).\eq{SGBT22}
\ee
Next, Let $u_{12}$ be the solution obtained by applying a BT with $a_2$
to $u_1$, and $u_{21}$ be obtained by appying a BT with $a_1$ to $u_2$: 
\be
\frac{1}{2}(u_{12}-u_1)_x&=&a_2\sin\left(\frac{u_{12}+u_1}{2}\right),\eq{SGBT111}\\
\frac{1}{2}(u_{21}-u_2)_x&=&a_1\sin\left(\frac{u_{21}+u_2}{2}\right).\eq{SGBT222}
\ee
According to the superposition principle, these two routes yield the same identical solution, i.e., $u_{12}=u_{21}$\cite{roshi}. To find $u_{12}$, we evaluate the following combination:
\be
&&\frac{1}{2}\left[(u_{12}-u_1)_x-(u_1-u_0)_x-(u_{21}-u_2)_x+(u_2-u_0)_x\right]\cr
&&=a_2\sin\left(\frac{u_{12}+u_1}{2}\right)-a_1\sin\left(\frac{u_1+u_0}{2}\right)\cr
&&-a_1\sin\left(\frac{u_{21}+u_2}{2}\right)+a_2\sin\left(\frac{u_2+u_0}{2}\right).\eq{u12}
\ee
By using $u_{12}=u_{21}$, the lhs of this equation vanishes{\color{black}. 
Then the rhs is simplified  as
\be
&&a_2\sin\left(\frac{u_{12}+u_1}{2}\right)-a_1\sin\left(\frac{u_{1}+u_0}{2}\right)\cr
&&-a_1\sin\left(\frac{u_{21}+u_2}{2}\right)+a_2\sin\left(\frac{u_2+u_0}{2}\right)\cr
&&=2\sin\left(\frac{u_{21}+u_1+u_2+u_0}{4}\right)\cr
&&\times\left(a_2\cos\left(\frac{u_{12}+u_1-u_2-u_0}{4}\right)-a_1\cos\left(\frac{-u_{21}-u_2+u_1+u_0}{4}\right)\right)\cr
&&=0.
\ee
Here
\be
&&a_2\cos\left(\frac{u_{12}+u_1-u_2-u_0}{4}\right)-a_1\cos\left(\frac{-u_{21}-u_2+u_1+u_0}{4}\right)\cr
&&=(a_2-a_1)\left(\cos\left(\frac{u_{12}-u_0}{4}\right)\cos\left(\frac{u_1-u_2}{4}\right)\right)\cr
&&-(a_1+a_2)\left(\sin\left(\frac{u_{12}-u_0}{4}\right)\sin\left(\frac{u_1-u_2}{4}\right)\right)\cr
&&=0.
\ee
Then it follows that
\be
\cot\left(\frac{u_{12}-u_0}{4}\right)=\frac{a_1+a_2}{a_2-a_1}\tan\left(\frac{u_1-u_2}{4}\right).
\ee
Solving this equation for $u_{12}$ leads to
\be
u_{12}=u_0+4\arccot\left(\frac{a_1+a_2}{a_2-a_1}\tan\left(\frac{u_1-u_2}{4}\right)\right).
\ee
We have thus obtained the 2-soliton solution. Note that we can not obtain the result without the nonlinear superposition principle of the B\"{a}cklund transformation.

\section{Conclusion and Discussion}
In this paper we reviewed powerful analytical tools for studying soliton equations. First, we reviewed the gauge-theoretical approach  based on the zero-curvature condition. By introducing the Lie-algebra valued gauge filed $A_{\mu}(t,x),(\mu=0,1)$, soliton equations are expressed through the zero-curvature equation, $F_{\mu\nu}=\partial_\mu A_\nu-\partial_\nu A_\mu+[A_\mu, A_\nu]=0$, which arises as the compatibility condition of a pair of linear equations (covariant derivatives of a wave function with respect to $t$ and $x$).

In the B\"{a}cklund transformation of various soliton equations, the compatibility condition of two first-order equations is likewise employed to derive soliton equations. This methodological similarity motivates a closer investigation into their shared features and differences. In the gauge-theoretical method, both the gauge fields and the wave function appear explicitly, and the soliton equations reside within the zero-curvature equation formed by these gauge fields. Conversely, in the B\"{a}cklund transformation, gauge fields do not appear; only the scalar/wave fields are present. This presents a fundamental conceptual difference between the two methods, despite both relying on compatibility conditions.

Soliton solutions are characterized by their soliton numbers, and establishing a systematic procedure to generate an $(N+1)$-soliton solution from an $N$-soliton solution is primary interest. For instance, in the stationary axisymmetric Einstein equation, solutions are known to be classified by soliton numbers,  and a corresponding B\"{a}cklund transformation has been established. Developing a similar recursive scheme for general soliton equations remains highly desirable.

Finally, we reviewed Hirota's bilinear method. The $\tau$-function formulation provides powerful framework for systematically consturcting multi-soliton solutions. If a general recursive relation connecting the $N$-soliton and $(N-1)$-soliton solutions is established, it will enable the generation of arbitrary $N$-soliton solutions starting from a trivial background.

\end{document}